\documentclass[times,doublespace]{stvrauth}

\usepackage{moreverb}

\usepackage{amsmath}
\usepackage{mathptmx}		

\usepackage[colorlinks,bookmarksopen,bookmarksnumbered,citecolor=red,urlcolor=red]{hyperref}

\newcommand\BibTeX{{\rmfamily B\kern-.05em \textsc{i\kern-.025em b}\kern-.08em
T\kern-.1667em\lower.7ex\hbox{E}\kern-.125emX}}

\usepackage{cite}
\begin{document}

\runningheads{M. Aghagholizadeh}{Study on Dynamics of Elastic Oscillator with a Rocking Wall}

\title{Study on Dynamics of an Elastic Oscillator Coupled with a Rocking Wall}

\author{Mehrdad Aghagholizadeh \corrauth}

\address{Department of Civil, Environmental and Construction Engineering, University of Central Florida, Orlando, FL 32816, USA}

\corraddr{Mehrdad Aghagholizadeh, Department of Civil, Environmental and Construction Engineering, University of Central Florida}

\begin{abstract}
This paper studies the dynamics of an elastic single degree of freedom oscillator (representing an elastic frame) coupled with a rocking wall. Two types of rocking walls namely stepping rocking wall and pinned rocking wall are presented and analyzed. For each case, full nonlinear equations of motions are calculated. The dynamic behavior of the systems shows mixed results in suppressing the dynamic response of the elastic oscillator. Through comprehensive analysis, pinned rocking wall amplifies the displacement along wide range of the spectrum, in the other hand, stepping rocking wall is the most effective especially in relatively flexible structures and with a heavier wall. This is mainly because of the pinned wall’s mass works against its stability. In this study, a simple, oscillator-rocking-wall model is defined and analyzed using OpenSees and, the results from OpenSees shows a good agreement with equation of motion solution using MATLAB.
\end{abstract}


\maketitle

\footnotetext[2]{Email: mehrdad@knights.ucf.edu}

\vspace{-6pt}

\section{Introduction}
\vspace{-2pt}
In the wake of severe damage to Olive View Hospital during the 1971 San Fernando, California earthquake, Bertero, et al., \cite{RN77}, directed attention of the engineers to coherent acceleration pulses. These pulses in the earthquake time history results large displacement demands in the structures. Also during 1994 Northridge, California and 1995 Kobe, Japan earthquakes many structures – especially tall moment resisting frames – which designed by that times seismic codes failed during the earthquake because of short story failure \cite{RN32, RN120, RN86, RN73}.

To prevent the soft story failure in structures, various studies had been conducted \cite{RN56, RN20, RN18}. One of the early works that introduced the concept of coupling a rocking wall with a moment resisting frame was work of \cite{RN114} and recently the seismic retrofitting of an 11-story building in Tokyo University in Japan has been done using a pinned rocking wall \cite{RN5, RN43}. Following the works of \cite{RN5, RN43}, several publications appeared to promote the seismic protection of a moment resisting frame structure when coupled with a rocking wall \cite{RN1, RN6, RN100, RN67}.  Also with the progress that has been made in the technology of precast shear walls as seismic resisting system for structures in seismically active areas (PCI Ad Hoc Committee on Precast Walls, \cite{RN112}) several studies conducted using precast walls \cite{RN29, RN8, RN7, RN63, RN62}. 

Most of the studies mentioned above are based on the seminal paper by Housner \cite{RN24}, which introduced advantages of rocking solitary column. Theses tall, slender columns exhibit remarkable performance and seismic stability. In his 1963 paper Housner shows that there is a safety margin between uplifting and overturning and that as the size of the free-standing column increases or the frequency of the excitation pulse increases, this safety margin increases appreciably to the extent that large free-standing columns enjoy ample seismic stability. Also \cite{RN65} recently explained that as the size of the free-standing rocking column increases, the enhanced seismic stability primarily originates from the difficulty to mobilize the rotational inertia of the column (wall) which increases with the square of the column (wall) size.

Accordingly, while, it becomes evident that most of the seismic resistance of tall free-standing columns (or walls) essentially originates from the difficulty to mobilize their large rotational inertia, the main emphasis on the behavior and capacity analysis of the coupled moment-frame-rocking-wall system as documented in the above-referenced studies is on the inelastic behavior of the structural system (inelastic behavior of the rocking wall-foundation interface) without analyzing the true dynamics of the system and the potential significance of considering the coupled dynamic effects. Clearly, there are cases where the response of the moment-resisting frame dominates the overall response and the rotational inertia effects of the rocking wall are negligible. Nevertheless, given that in principle the dynamics of the rocking wall is not negligible and in some cases, it may be unfavorable since it may drive the structure; the main motivation for this study is to examine to what extent the dynamics of a stepping or a pinned rocking wall influence the dynamic response of the coupled elastic oscillator.

The motivation for coupling of a moment-resisting frame with a strong rocking wall is to primarily enforce a uniform distribution of interstory drifts; therefore, the first mode of the frame becomes dominant as was first indicated in the seminal paper by Alavi and Krawinkler \cite{RN86}. Further analytical evidence to the first-mode dominated response is offered in \cite{RN5}. These results together with additional evidence by other investigators were critically evaluated in a recent paper Grigorian \cite{RN1}, who concluded that a moment resisting frame coupled with a rocking wall can be categorized as a single-degree-of-freedom (SDOF) system. Accordingly, in this study we adopted the SDOF idealization shown in Figure 1.

\vspace{-6pt}

\section{DYNAMICS OF AN ELASTIC OSCILLATOR COUPLED WITH A STEPPING ROCKING WALL}

Dynamics of an elastic single degree of freedom oscillator coupled with a stepping rocking wall is investigated in this part. The schematic of the problem is shown in Figure 1. An oscillator with stiffness k, damping of c and the mass of $m_w$ is coupled with a stepping rocking wall with mass of $m_w$, wall size of $R=\sqrt{b^2+h^2 }$, slenderness, $tan\alpha=b⁄h$ and moment of inertia about pivoting points $O$ and $O'$, $I=4/3 m_w R^2$.  For the sake of simplicity, it is assumed that the link between the wall and the oscillator is located between center of the mass of the wall and oscillator in the height of h from the foundation of the stepping rocking wall as shown in Figure 1.
\begin{figure}[t]
\centering
\includegraphics[width=10cm]{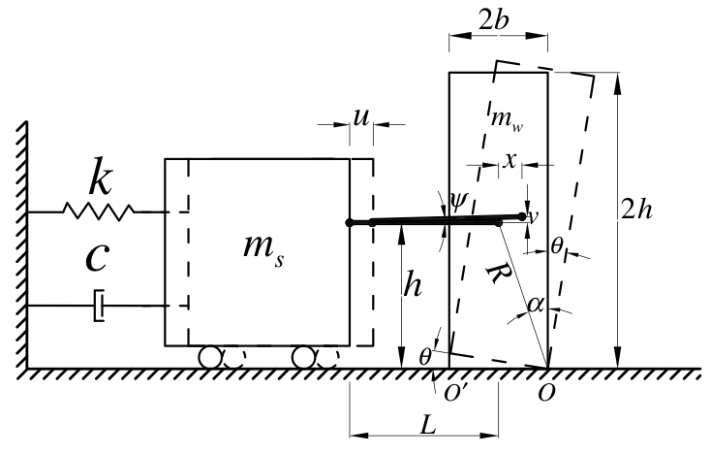}
\caption{Elastic SDOF oscillator coupled with a stepping rocking wall}
\label{fig:1}
\end{figure}
While the block starts to uplift, center of mass of the wall goes upward by $v$, so the coupling arm rotates by an angle of $\phi$. So, the translation of center of mass of the wall, $x$, is related to horizontal displacement of the oscillator mass, $m_s$, and can be expressed via, $cos\phi=1-(u-x)/L$; in which $\phi=sin^{-1}⁡(v/L)$. Hence the horizontal displacement, $u$, is related to the horizontal displacement of center of the mass of the wall,$x$, through the following equation:

\begin{equation} \label{eq:1a}
\dfrac{u}{L}=1+\dfrac{x}{L}-\sqrt{1-\dfrac{v^2}{L^2}}
\end{equation}

In this paper, the coupling arm is assumed to be long enough so that $v^2/L^2$ is much smaller that unity $(v^2⁄L^2 <<1)$; and in this case $u=x$. Clearly, there are cases where the coupling arm is short and in this case the term $v^2/L^2$ is not negligible. Nevertheless, a recent study by Makris and Aghagholizadeh 2016 \cite{RN119} on the response of an elastic oscillator coupled with a rocking wall showed that the effect due to a shorter coupling arm is negligible.

The system under consideration is a single-degree-of-freedom system where the lateral translation of the mass, u is related to the rotation of the stepping rocking wall $\theta$ via the expression:
\begin{equation} \label{eq:1}
u=\pm R[\sin\alpha-\sin(\alpha \mp \theta)]
\end{equation}
\begin{equation} \label{eq:2}
\dot{u}= R~\dot{\theta} \cos(\alpha \mp \theta)
\end{equation}
\begin{equation} \label{eq:3}
\ddot{u}=R[\ddot{\theta}\cos(\alpha \mp \theta) \pm \dot{\theta^2}\sin(\alpha \mp \theta)    ]
\end{equation}
In equations (\ref{eq:1}) to (\ref{eq:3}) whenever there is a double sign (say $\pm$), the top sign is for $\theta>0$ and the bottom sign is for $\theta<0$.

\noindent Dynamic equilibrium of the mass $m_{s}$ gives:
\begin{equation} \label{eq:4}
m_{s} (\ddot{u}+\ddot{u}_{g})=-ku-c \dot{u}+T
\end{equation}
In equation (\ref{eq:4}), $T$, represents the axial force in the coupling arm which is positive when force is positive.\\

\noindent \textit{Case 1: $\theta>0$} :

\noindent For positive rotations $(\theta>0)$, dynamic equilibrium of the roting restrained stepping wall with mass $m_w$, gives:
\begin{equation} \label{eq:7}
\begin{split}
I\ddot{\theta}=-TR\cos(\alpha-\theta)-m_w g R\sin(\alpha-\theta)-m_w \ddot{u}_g R\cos(\alpha-\theta)
\end{split}
\end{equation}
\noindent The axial force $T$ appearing in equation (\ref{eq:7}) is replaced with the help of equation (\ref{eq:4}) and for a rectangular stepping wall $(I=\dfrac{4}{3}m_w R^2$  ), equation (\ref{eq:7}) assumes the form:
\begin{equation} \label{eq:8}
\begin{aligned}
&\dfrac{4}{3} m_w R^2 \ddot{\theta}+[m_s (\ddot{u}+\ddot{u}_g)+
ku+c\dot{u} ]R \cos(\alpha-\theta)=-m_w R[\ddot{u}_g\cos(\alpha-\theta)
+g \sin(\alpha-\theta) ]
\end{aligned}
\end{equation}

\noindent upon dividing with $m_wR^2$ and inserting equations (\ref{eq:1}) to (\ref{eq:3}) instead of  $u$, $\dot{u}$ and $\ddot{u}$,  equation (\ref{eq:8}) assumes the form:
\begin{equation}  \label{eq:9}
\begin{aligned}
&[\frac{4}{3}+\gamma \cos^2⁡(\alpha-\theta) ]\ddot{\theta}+\gamma \cos⁡(\alpha-\theta) \big[\omega_o^2 (\sin⁡\alpha-sin⁡(\alpha-\theta) ) \\
&+2\xi\omega_o \dot{\theta} \cos⁡(\alpha-\theta)+\dot{\theta}^2  sin⁡(\alpha-\theta) \big]=-\frac{g}{R} \big[(\gamma+1)\frac{\ddot{u}_g}{g} \cos⁡(\alpha-\theta)+\sin⁡(\alpha-\theta)\big]
\end{aligned}
\end{equation}

\noindent in which $\gamma=m_s/m_w$ is the mass ratio parameter, $\omega_o=\sqrt{k/m_s}$ is undamped frequency and $\xi$ is the viscous damping ratio of the SDOF oscillator.\\

\noindent \textit{Case 2: $\theta<0$}:\\
\noindent For negative rotations one can follow the same reasoning and the equation of the coupled system shown in Figure (\ref{fig:1}) is:
\begin{equation}  \label{eq:12}
\begin{aligned}
&[\frac{4}{3}+\gamma \cos^2⁡(\alpha+\theta) ]\ddot{\theta}-\gamma \cos⁡(\alpha+\theta) \big[\omega_o^2 (\sin⁡\alpha-sin⁡(\alpha+\theta) ) \\
&-2\xi\omega_o \dot{\theta} \cos⁡(\alpha+\theta)+\dot{\theta}^2  sin⁡(\alpha+\theta) \big]=\frac{g}{R} \big[-(\gamma+1)\frac{\ddot{u}_g}{g} \cos⁡(\alpha+\theta)+\sin⁡(\alpha+\theta)\big]
\end{aligned}
\end{equation}
In equations (\ref{eq:9}) and (\ref{eq:12}), term that are multiplied by $\gamma=m_s⁄m_w$ , are related to dynamic response of elastic oscillator and other terms are related to dynamics of stepping rocking wall. In the absence of the elastic oscillator ($\gamma=\omega_o=\xi=0$), equations (\ref{eq:9}) and (\ref{eq:12}) reduce to the equation of motion of the solitary free-standing column \cite{RN79, RN81}.

\noindent During the oscillatory motion of the coupled system shown in Figure~\ref{fig:1}, aside from the energy that is dissipated from the inelastic behavior of the SDOF oscillator and the idealized viscous damping, additional energy is also lost during impact when the angle of rotation reverses. At this instant it is assumed that the rotation of the rocking wall continues smoothly from points $O$ to $O'$; nevertheless, the angular velocity, $\dot{\theta}_2$, after the impact is smaller than the angular velocity, $\dot{\theta}_1$, before the impact. Given that the energy loss during impact is a function of the wall-foundation interface, the coefficient of restitution, $e=\dot{\theta}_2/\dot{\theta}_1<1$, is introduced as a parameter of the problem. In this study the coefficient of restitution assumes the value of $e= 0.9$.\\

Minimum acceleration needed to initiate rocking can be calculated as follows: \cite{RN165, RN119, RN166}.
\begin{equation}  \label{eq:13}
\ddot{u}_g \geqslant \dfrac{g~tan\alpha}{\gamma+1}
\end{equation}

\section{DYNAMICS OF AN ELASTIC OSCILLATOR COUPLED WITH A PINNED ROCKING WALL}
Using frame retrofitting method that introduced by \cite{RN86}, in their study \cite{RN43} and \cite{RN5} proposed a pinned rocking wall for the seismic protection of an 11-story moment resistant frame in Tokyo University, Japan. The novelty in these studies is that the rocking wall does not alternate pivot points (it is not a stepping wall) given that it is pinned at mid-width as shown in Figure 2.
\begin{figure}[b]
\centering
\includegraphics[width=10cm]{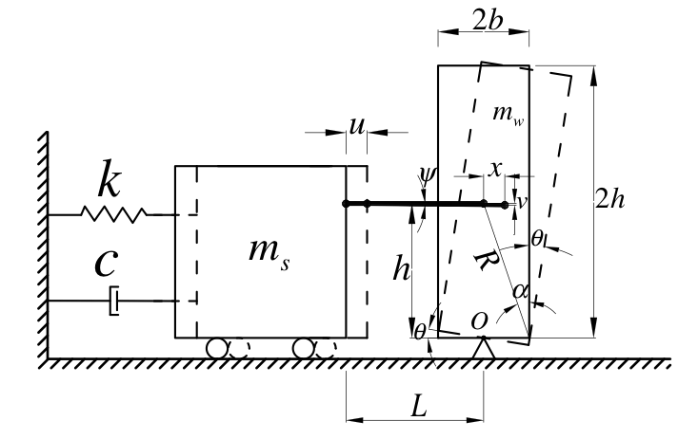}
\caption{Elastic SDOF oscillator coupled with a pinned rocking wall}
\label{fig:2}
\end{figure}
The pinned rocking wall shown in Figure 2 is a SDOF (with the same reasoning of the previous case) system and translation of oscillator mass can be expressed in terms of rotation of the pinned wall and can be expressed as follows:

\begin{equation}
u=h\sin\theta
\label{eq:pin_disp}
\end{equation}
And time derivations can be expressed as:
\begin{equation}
\dot{u}=h\dot{\theta}\cos\theta
\label{eq:pin_velo}
\end{equation}
\begin{equation}
\ddot{u}=h\ddot{\theta}\cos\theta-h\dot{\theta}^2\sin\theta
\label{eq:pin_acc}
\end{equation}
System shown in Figure 2 is a single degree of freedom oscillator with mass, $m_s$, stiffness, k, and damping c, that is coupled with pinned wall of size $R=\sqrt{b^2+h^2}$, slenderness, $tan⁡\alpha=b/h$, mass, $m_w$ and moment of inertia about the pin O, $I=m_w R^2 (1/3+cos^2\alpha)$.
Dynamic equilibrium of the mass, $m_s$, of the oscillator is similar to equation 5. In this case equation of motion for the pinned rocking wall is the same for positive and negative rotation:
\begin{equation}
I\ddot{\theta}=-Th \cos⁡\theta+m_w gh sin⁡⁡\theta-m_w \ddot{u}_g h cos⁡⁡\theta
\label{eq:15_p}
\end{equation}
Note that as it can be seen in equation (\ref{eq:15_p}), wall mass $m_w$, in this case works against the stability of the system. With similar steps as it described for the stepping rocking wall one can derive the equation of motion for pinned rocking wall using equations (\ref{eq:pin_disp}) to (\ref{eq:15_p}). 
\begin{equation}
\begin{aligned}
&[\frac{1}{3}+(1+\gamma \cos^2⁡\theta)cos^2⁡\alpha]\ddot{\theta}+\gamma \cos^2\alpha \cos⁡⁡\theta [(\omega_o^2-⁡\dot{\theta}^2 ) \sin⁡\theta+2\xi \omega_o \dot{\theta}cos⁡⁡\theta]\\
&=-\frac{g}{R}  cos⁡\alpha [(\gamma+1) \frac{\ddot{u}_g}{g}  cos\theta-sin⁡⁡\theta]
\end{aligned}
\label{eq:EOM_pin}
\end{equation}
Equation (\ref{eq:EOM_pin}) is equation of motion for pinned rocking wall both for positive and negative rotations and all the parameters are similar to the stepping rocking wall.

\section{RESPONSE SPECTRA OF AN ELASTIC OSCILLATOR COUPLED WITH A ROCKING-WALL}
\begin{figure}[t]
\centering
\includegraphics[width=11cm]{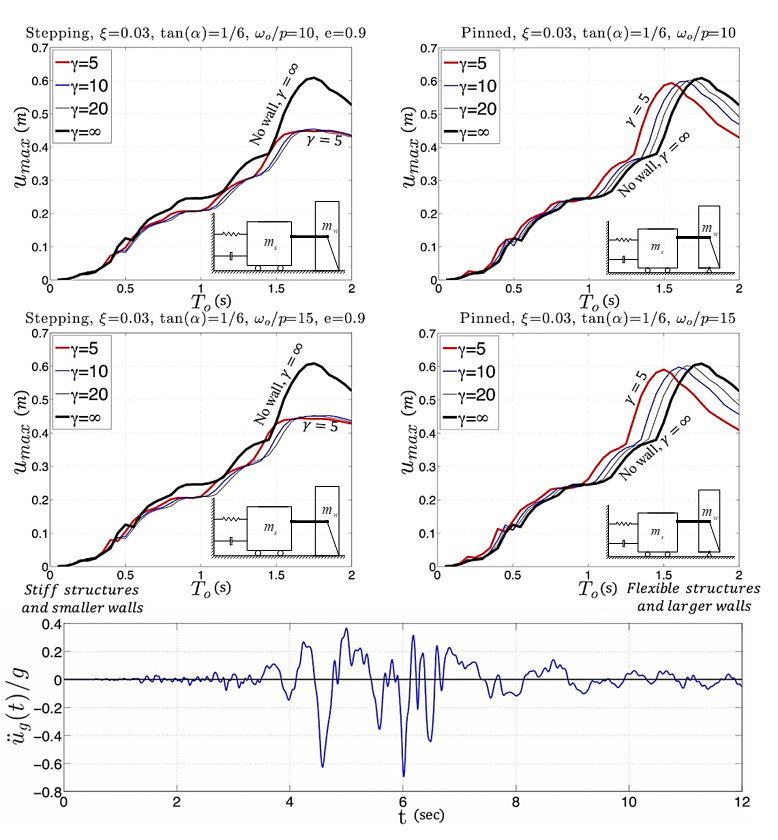}
\caption{Displacement spectra of an elastic SDOF oscillator coupled with a stepping wall (left) and a pinned wall (right) for three values of the mass ratio $\gamma=m_s/m_w=5$,10 and 20 and two values of the wall size, $\omega_o/p=10$ and 15 when subjected to the Takarazuka/000 ground motion recorded during the 1995 Kobe, Japan earthquake (bottom).}
\label{fig:kobe}
\end{figure}
In order to find earthquake response spectra of the systems shown in Figures 1 and 2, equations (\ref{eq:9}), (\ref{eq:12}) and (\ref{eq:EOM_pin}) are used. In Figure 3 displacement spectra for stepping rocking wall (left) versus pinned rocking wall (right) is shown when systems are excited by the Takarazuka/000 ground motion recorded during the 1995 Kobe, Japan earthquake (bottom). The top plots are for $\omega_o/p=10$; whereas the bottom plots are for $\omega_o/p=15$---that is for a larger wall at any given structural frequency, $\omega_o=2\pi/T_o$.

\begin{figure}[t]
\centering
\includegraphics[width=11cm]{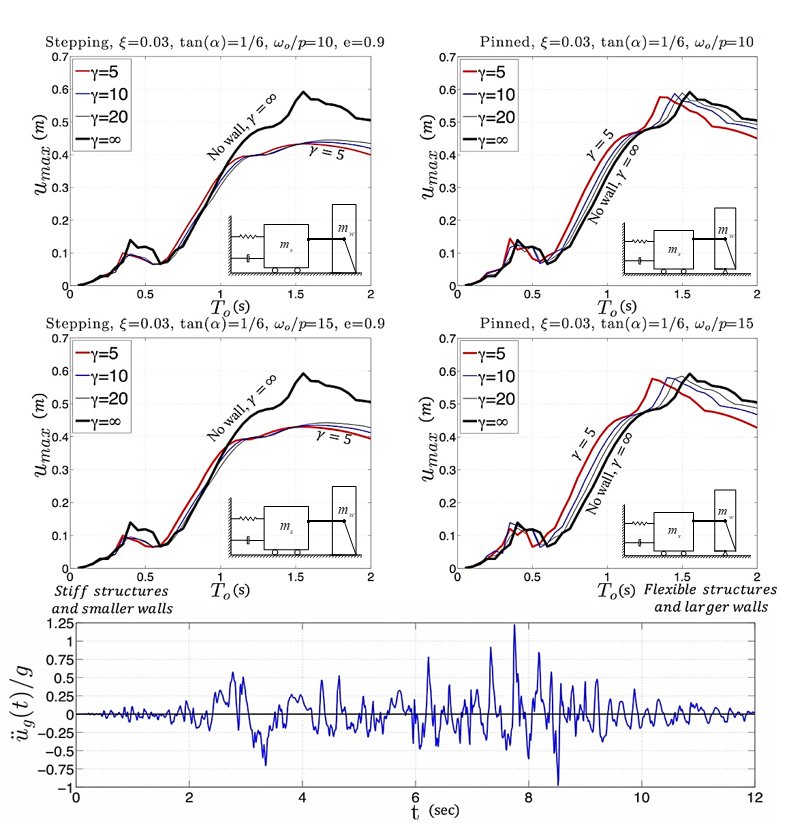}
\caption{Displacement spectra of an elastic SDOF oscillator coupled with a stepping wall (left) and a pinned wall (right) for three values of the mass ratio $\gamma=m_s/m_w=5$,10 and 20 and two values of the wall size, $\omega_o/p=10$ and 15 when subjected to the Pacoima Dam/164 ground motion recorded during the 1971 San Fernando, California earthquake (bottom).}
\label{fig:PCD}
\end{figure}

When reading the earthquake spectra shown in Figures \ref{fig:kobe} and \ref{fig:PCD} the reader needs to recognize that as the period, $T_o$ of the SDOF oscillator increases, for a given ratio of $\omega_o/p$, the size of the coupled wall also increases. For instance, for the top plots which are for $ω_o/p=10$, the frequency parameter, p, of the wall that is coupled to a structure with $T_o=0.5$ sec is $p=\omega_o/10=\frac{2\pi}{0.5}\frac{1}{10}=1.26$ rad/sec, which corresponds to a value of $R=3g/4p^2=4.66$ m; therefore, the wall with slenderness, $tan⁡\alpha=1/6$, is $9.20$ m tall.

When a structure with $T_o=1.0$ sec is of interest, the frequency parameter, p, of the wall is $p=\omega_o/10=\frac{2\pi}{10}\frac{1}{10}$ rad/sec, which corresponds to a value of $R=3g/4p^2=18.6$ m; therefore, the wall with a slenderness, $tan⁡\alpha=1/6$,  is 36.80 m tall. When observing Figure (\ref{fig:kobe}), what is worth noting is that in the case where the SDOF oscillator is coupled with a stepping wall (left plots), the presence of the stepping wall suppresses the displacement response (with the heavier wall, $\gamma=5$ being most effective), for flexible structures (large values of $T_o$).
In contrast in the case where the SDOF oscillator is coupled with a pinned wall (right plots), the presence of the pinned wall amplifies the response for the most of the spectrum with the heavier wall ($\gamma=5$) being most detrimental. This mainly happens because in the case of the pinned wall, the moment from its weight $=+m_w gh sin⁡\theta$ works against stability as shown in equation (15). 

Similar trends are shown in Figure 4 which shows displacement spectra when the coupled elastic SDOF oscillator-rocking wall system is subjected to the Pacoima Dam/164 ground motion recorded during the 1971 San Fernando, California earthquake. (In addition to earthquake spectra \cite{RN119}, analyzed spectra of these systems under symmetric Ricker wavelet \cite{RN96} pulse acceleration).
\section{OpenSees MODELING OF AN ELASTIC OSCILLATOR COUPLED WITH A STEPPING ROCKING WALL}
\begin{figure}[t]
\centering
\includegraphics[width=9cm]{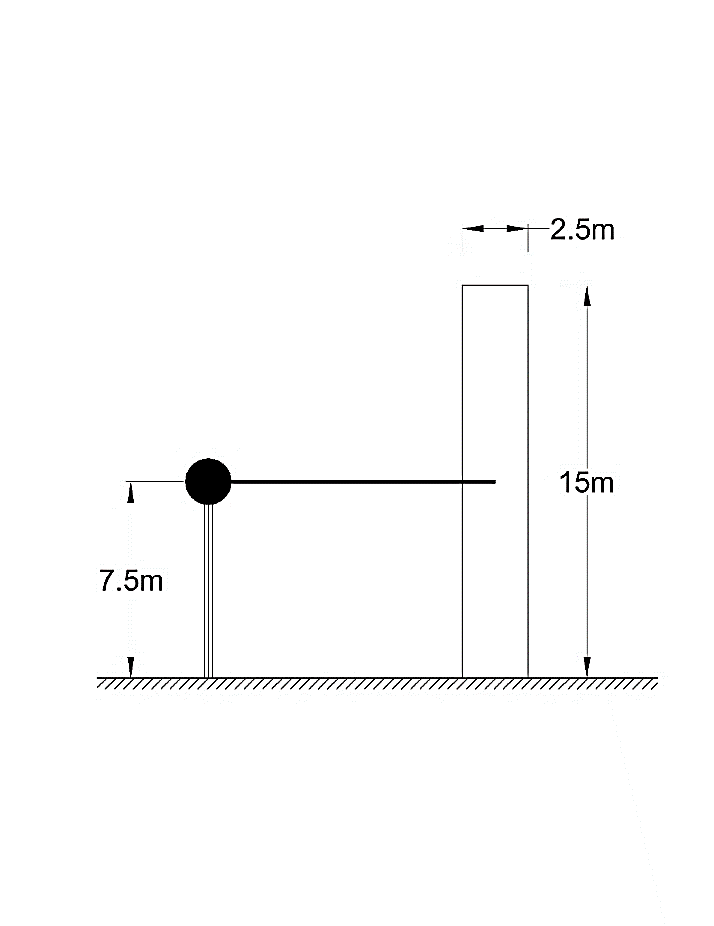}
\caption{Simple OpenSees model representing an elastic oscillator coupled with a stepping rocking wall}
\label{fig:Ops}
\end{figure}

In this section, a simple model representing an elastic oscillator coupled with a stepping rocking wall is presented. The system is shown in Figure (\ref{fig:Ops}) is a fixed end column with period of $T_o=0.64$ s, and a concentrated mass at the top, $m_s$. The column model defined using elastic beam column element in OpenSees \cite{RN59}. The rocking surface between ground and bottom of the stepping rocking wall is modeled using zero-length fiber cross section element with nonlinear elastic compression and no tension material, placed between them. This type of cross section enables simulation of the rocking motion.
The only issue with this type of model is the energy dissipation of the wall when it changes the pivot point cannot be considered. To simulate energy dissipation in each impact, a rotational viscous damper is defined at the bottom of the wall. The specification of this damper and its coefficient is selected using study of \cite{RN22}.  Viscous damper constant is defined as follows:
\begin{equation}
c=110~\alpha^2m_wg^{0.5}R^{1.5}
\end{equation}
In which $\alpha$ is the wall slenderness, $m_w$, is the wall mass and R, is wall size (as all shown in equations of motion calculated in previous sections).
\begin{figure}[b]
\centering
\includegraphics[width=12cm]{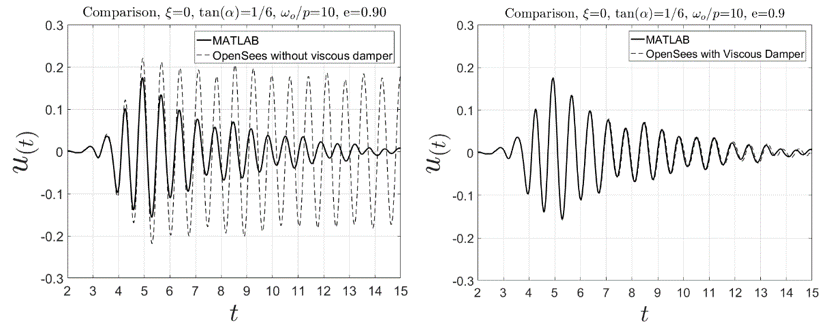}
\caption{Response of the system (with) and without (left) damper) when subjected to the CO2/065 ground motion recorded during the 1966 Parkfield, California earthquake.}
\label{fig:Comp}
\end{figure}
Figure (\ref{fig:Comp}) shows response of the system when subjected to the $CO2/065$ ground motion recorded during the 1966 Parkfield, California earthquake. Responses of the system—shown in Figure (\ref{fig:Ops})---using OpenSees framework is compared with the results of a system with similar parameters (period $T_o=0.64 s$ and mass ration, $\gamma=m_s⁄m_w=5$) using equation 8 and 9 from MATLAB. Figure (\ref{fig:Comp}-left) shows response of the system when there is no viscous damper added. Similarly, response of the system with viscous damper is shown in Figure (\ref{fig:Comp}-right).  

This result clearly shows that using a rotational viscous damper is a practical way to simulate energy dissipation during wall impact and the results have a good agreement with the solution from equations of motion 8 and 9.

\section{Conclusion}
This paper studied dynamics of a single-degree-of-freedom, elastic oscillator when it is coupled with a stepping rocking wall and pinned rocking wall. The full nonlinear equations of motion for both cases have been derived and analyzed subjected to different earthquake time histories. This study reaches to the following conclusions.

In the case that SDOF oscillator is coupled with a stepping rocking wall, presence of the wall suppresses the displacement of the system, especially for flexible oscillators. In the other hand, pinned rocking wall amplifies the responses of the system, and heavier wall has more amplification disadvantage. This happens mainly because the weight of pinned wall works against the stability of the system.

Also, a simple model of oscillator coupled with a stepping rocking wall is modeled and analyzed using OpenSees framework. This study showed that in order to capture the energy dissipation of the wall when it changes the pivot point, using a rotational viscous damper is a practical an accurate method. Comparison of time history response of the system compared with equation of motion solution from MATLAB shows a good agreement.

The study of the response of a yielding SDOF oscillator coupled with a rocking wall is ongoing and will be presented in a future publication.

\ack The author wishes to express his acknowledgement and gratitude to Dr. Nicos Makris whose guidance and comments helped throughout this research.

\end{document}